\documentclass[final,1p,times]{elsarticle} 
\usepackage{graphicx} 
\usepackage{amssymb} 
\usepackage{amsthm} 
\usepackage{lineno} 

\journal{Nuclear Physics A} 
\begin{document} 

\begin{frontmatter} 


\title{Dilepton radiation measured in PHENIX probing the strongly
       interacting matter created at RHIC}

\author{Yasuyuki Akiba$^{a}$ for the PHENIX collaboration
}

\address[a]{RIKEN Nishina Center for Accelerator-Based Science, 
Hirosawa, Wako, Saitama 351-0198, Japan}

\begin{abstract} 
PHENIX has measured $e^+e^-$ pairs from $p+p$ and Au+Au collisions as function of
mass and $p_T$. The data can be used to probe the properties of dense matter formed
in Au+Au collision. The relation between electron pairs and virtual photons is discussed.
\end{abstract} 

\end{frontmatter} 



\section{Introduction}
It is now well established that high density partonic matter is formed in collisions
of heavy nuclei at RHIC\cite{PHENIX_WP}.
Photons and lepton pairs are the cleanest probes of the dense matter.
Since these electromagnetic probes have little interaction with the matter, they carry direct
information deep inside of the medium. From the emission
rate of photons and dileptons, we can directly probe the temperature of the medium,
the properties of hadrons inside of the medium, and the matter's properties.

\section{Relation between dilepton and virtual photon}
The thermal emission rate of electron pairs per space-time volume can be
described in terms of the electromagnetic (EM) spectral function as \cite{rapp0}
\begin{eqnarray} \label{eq:spectral_fc}
\frac{dR_{ee}}{d^4q} = -\frac{\alpha^2}{3\pi^3}\frac{L(M)}{M^2}{\rm Im}\Pi^{\mu}_{em,\mu}(M,q;T)f^B(q_0,T),\\
L(M)=\sqrt{1-\frac{4m_e^2}{M^2}}(1+\frac{2m_e^2}{M^2}).
\end{eqnarray}
Here $\Pi^{\mu}_{em,\nu}$ is the in-medium EM spectral function
and $f^B(q_0,T)=1/(e^{q_0/T}-1)$ is the Boltzmann factor.
The equation shows that from the emission rate we can probe the medium property
encoded in the EM spectral function as well as its temperature in the Boltzmann factor.
The EM spectral function in vacuum for the
low mass region is well described as a sum of contributions of the low mass vector
mesons ($\rho$,$\omega$, and $\phi$). Modification of the properties of these mesons shows up
as modification of the EM spectral function.

Using the same notation, the emission rate of virtual photons is
described as \cite{rapp0,Turbide2008}
\begin{eqnarray}\label{eq:vphoton}
q_0\frac{dR_{\gamma^*}}{d^3q} = -\frac{\alpha}{2\pi^2}{\rm Im}\Pi^{\mu}_{em,\mu}(M,q;T)f^B(q_0,T).
\end{eqnarray}
The virtual photon and the electron pair emission rate are related as
\begin{eqnarray}
q_0\frac{dR_{ee}}{dM^2d^3q} = \frac{1}{2} \frac{dR_{ee}}{d^4q}
= \frac{\alpha}{3\pi}\frac{L(M)}{M^2}q_0\frac{dR_{\gamma^*}}{d^3q}.\label{eq:virtual}
\end{eqnarray}
The same relation holds for the
yield of photons and electron pairs after space-time integration.
This relation between the virtual photon and electron 
pair emission is exact to the order of $\alpha$ in QED and is exact to all
orders of strong couplings.
For $M\rightarrow 0$, the virtual photon yield becomes the real photon yield
($N_{\gamma^*} \rightarrow N_{\gamma}$). Thus this relation can be used to 
determine the yield of real photons. Recently, PHENIX measured the yield of 
direct photon in Au+Au from the yield of low mass $e^+e^-$ pairs
using this relation.\cite{ppg086}.

The equation (\ref{eq:virtual}) can be rewritten as
\begin{eqnarray}
q_0\frac{dN_{\gamma^*}}{d^3q} 
\simeq \frac{3\pi}{2\alpha} M q_0\frac{dN_{ee}}{d^3qdM}.\label{eq:lepton_photon}
\end{eqnarray}
This equation can be used to convert the yield of $e^+e^-$ pair to virtual
photon yield. Such conversion is useful since the virtual photon emission rate
is just the product of the EM spectral function and the Boltzmann factor.

\begin{figure}[ht]
\centering
\includegraphics[width=0.4\linewidth]{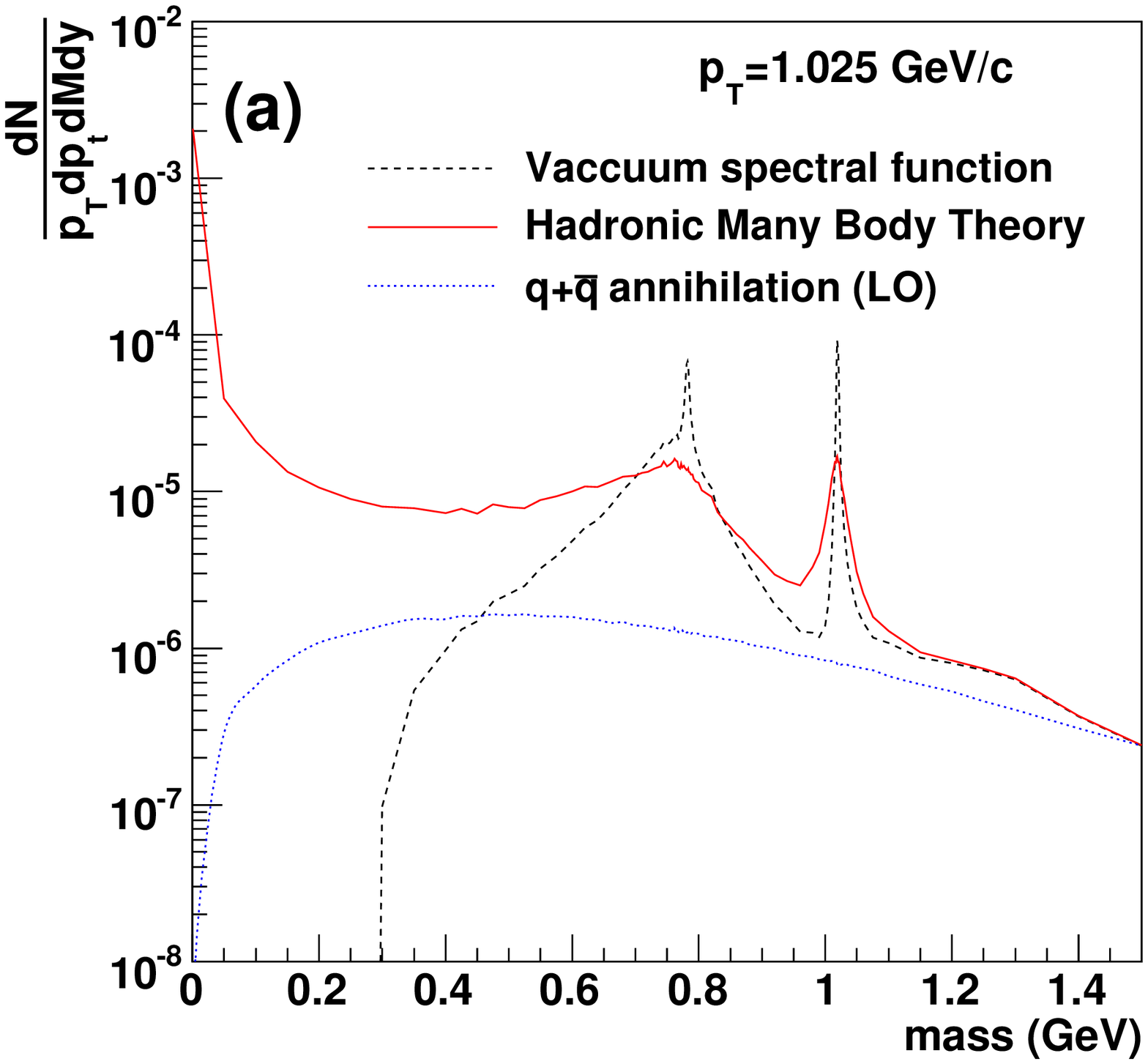}
\includegraphics[width=0.4\linewidth]{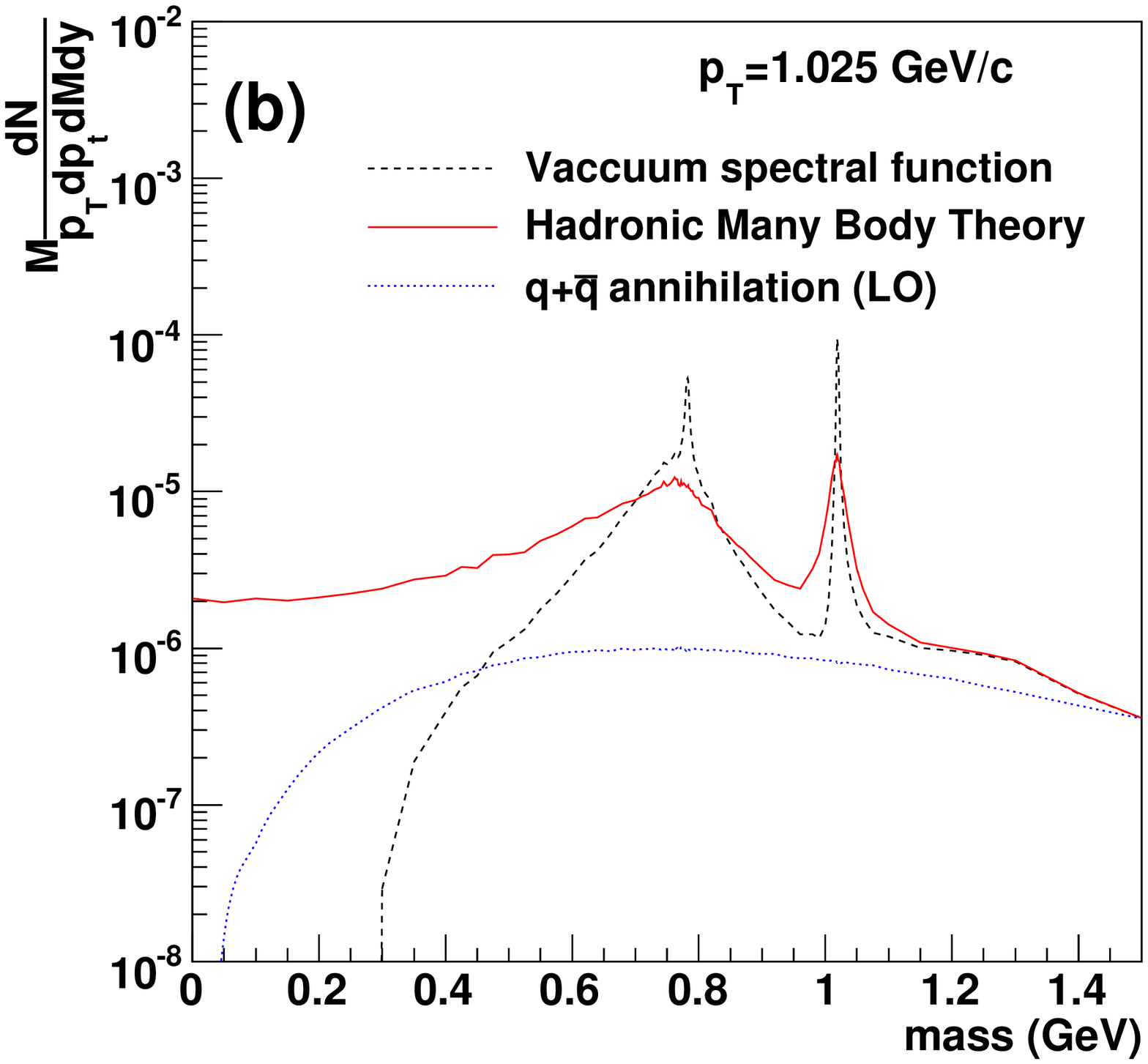}
\caption[]{
\label{fig:RalfRapp}
Electron pair emission rate (left panel) and virtual photon emission rate
(right panel) calculation by R.~Rapp\cite{rapp1}. The solid and dashed curve uses
the EM spectral function in the medium and in vacuum, respectively. 
The dotted curve shows $q\bar{q}$ annihilation contribution.  }
\end{figure}

Figure~\ref{fig:RalfRapp} illustrates the relation between electron pair mass spectrum 
and virtual photon yield.
The left panel of the figure shows the differential electron pair spectrum,
$(1/p_T)dN_{ee}/dMdp_Tdy$ at $p_T=1.025$ GeV/$c$ from a model calculation of electron pair
production by Rapp. The dashed and solid curve show electron pairs from hadronic
gas, while the dotted curve show the contribution from the leading order (LO) $q+\bar{q}$
corrected with a Hard Thermal Loop (HTL)
correction. The dashed curve uses
the EM spectral function $\Pi_{EM}$ that is unchanged from its vacuum value so the line shapes
of vector mesons ($\rho, \omega, \phi$) are unmodified. In the solid curve
the spectral function is calculated by hadronic many body theory (HMT), and
the vector mesons are broadened due to the interactions.
It also includes the contributions like
$a_1$(1260)$\rightarrow \pi+e^+e^-$,
$\rho \rightarrow \pi + e^+e^-$,
and $N+\pi \rightarrow N^* \rightarrow Ne^+e^-$. These contributions filled the low mass
regions below the two-pion threshold. In the low mass region, the mass spectrum steeply increases
with decreasing $M$. This behavior is due to the $1/M$ factor in 
$\gamma^* \rightarrow e^+e^-$.

The right panel of Fig.~\ref{fig:RalfRapp} shows the same calculations presented as
the yield of virtual photon. The steep $1/M$
behavior of the electron pair spectrum is removed, and much more
smooth behavior of the virtual photon spectrum is revealed. The plot
shows that the virtual photon yield is almost constant as a function of
$M$. The value of the solid curve at $M=0$ should correspond to the real
photon yield.
This illustrates that in a consistent theory calculation the
yield of virtual photon is a smooth function of $M$ and it becomes the
real photon yield in the limit of $M=0$.


The quark annihilation contribution, shown as dotted curve,
behaves as $\propto M^2$ in the right panel at
low mass region since $q\bar{q}$ contribution to $\Pi_{EM}$ is
proportional to $M^2$. Thus it is strongly suppressed and have little
contribution in the low mass region. In the high mass region, the
$M^2$ behavior of the quark annihilation is suppressed by the
Boltzmann factor.

It should be noted that the dotted curve does not include processes like
$q+g \rightarrow q+\gamma^*$ that are associated with real direct
photon production in QGP. This is because HTL calculation of
thermal radiation from QGP is only available in the real photon
case.
Turbide, Gale, and Rapp \cite{Turbide:2003si} calculated real photon
production in an hadronic gas using the same model and compared
it with real  photon production in QGP phase using the complete
leading order HTL analysis. They found that real photon from
the QGP outshines that of hadronic gas
for $p_T>1.5$ GeV/$c$ in Au+Au collisions at RHIC. This means that
contribution from processes associated with real photon
production in QGP can be as large as that of HMT (solid curve) and can be
much larger than that of the LO $q\bar{q}$ annihilation (dotted).

\begin{figure}
\includegraphics[width=\linewidth]{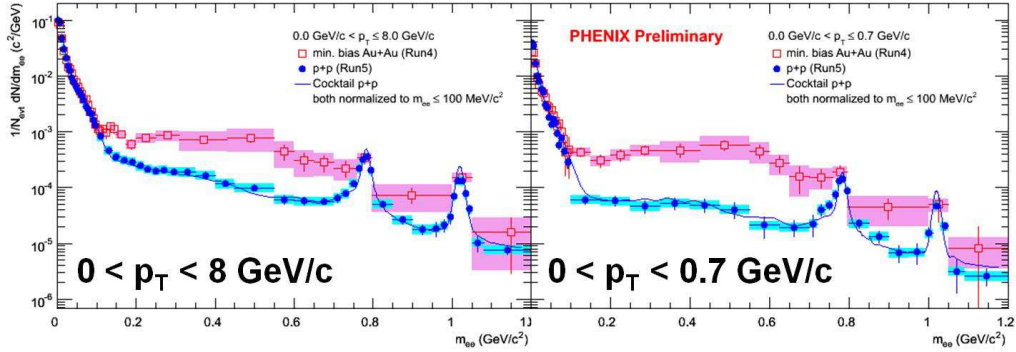}
\caption[]{
\label{fig2}
Mass distributions of $e^+e^-$ pair in $p+p$ (filled) and Au+Au (open). 
The left panel show the inclusive mass  spectra ($0<p_T<8$ GeV/$c$)
and the right panel shows the mass spectrum at the low $p_T$
region $(0<p_T<0.7)$ GeV/$c$.}
\end{figure}

\section{$e^+e^-$ mass spectra in $p+p$ and Au+Au}
PHENIX measured the $e^+e^-$ pair production in Au+Au and in $p+p$ at
$\sqrt{s_{NN}}$=200 GeV\cite{ppg075,ppg085}.
In $p+p$ the measured mass spectrum in low mass ($M<1$ GeV/$c^2$) is well described by 
the sum of light hadron decay contributions. The high-mass region is dominated
by the contribution of the correlated decays of charm and bottom. From the
measured mass spectrum the charm cross section is determined as
$\sigma_{c\bar{c}} = 544 \pm 39 \pm 142 \mu$b, which is consistent with that
obtained from single electron measurement\cite{ppg065}.
The bottom cross section is determined as $\sigma_{b\bar{b}} = 3.9\pm2.5^{+3}_{-2} \mu$b,
which is consistent with that obtained from $e-h$ correlation\cite{ppg094}.

Figure~\ref{fig2} compares the $e^+e^-$ pair mass distributions in $p+p$ and $Au+Au$ collisions
and hadronic cocktail for $M<1.2$ GeV/$c^2$. The $p+p$ and Au+Au data are normalized in
Dalitz pair mass region ($M<30$ MeV/$c^2$). While the $p+p$ data is well described by the
cocktail shown in solid curve, the Au+Au data show a large enhancement over the cocktail.
The enhancement is larger in the low $p_T$ region ($p_T<0.7$ GeV/$c$) shown in the left panel
of the figure.

\begin{figure}
\centering
\includegraphics[width=0.45\linewidth]{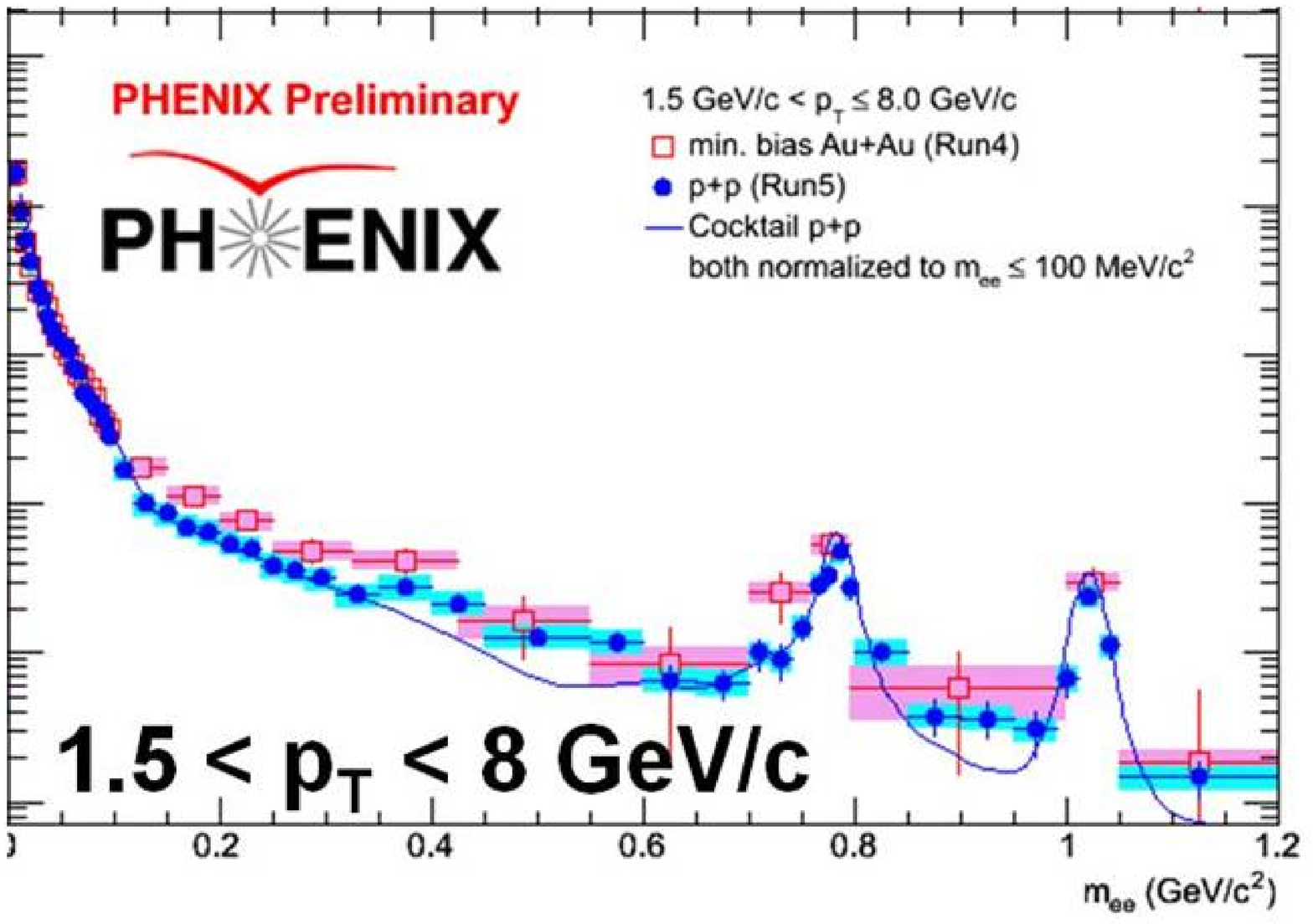}
\includegraphics[width=0.45\linewidth]{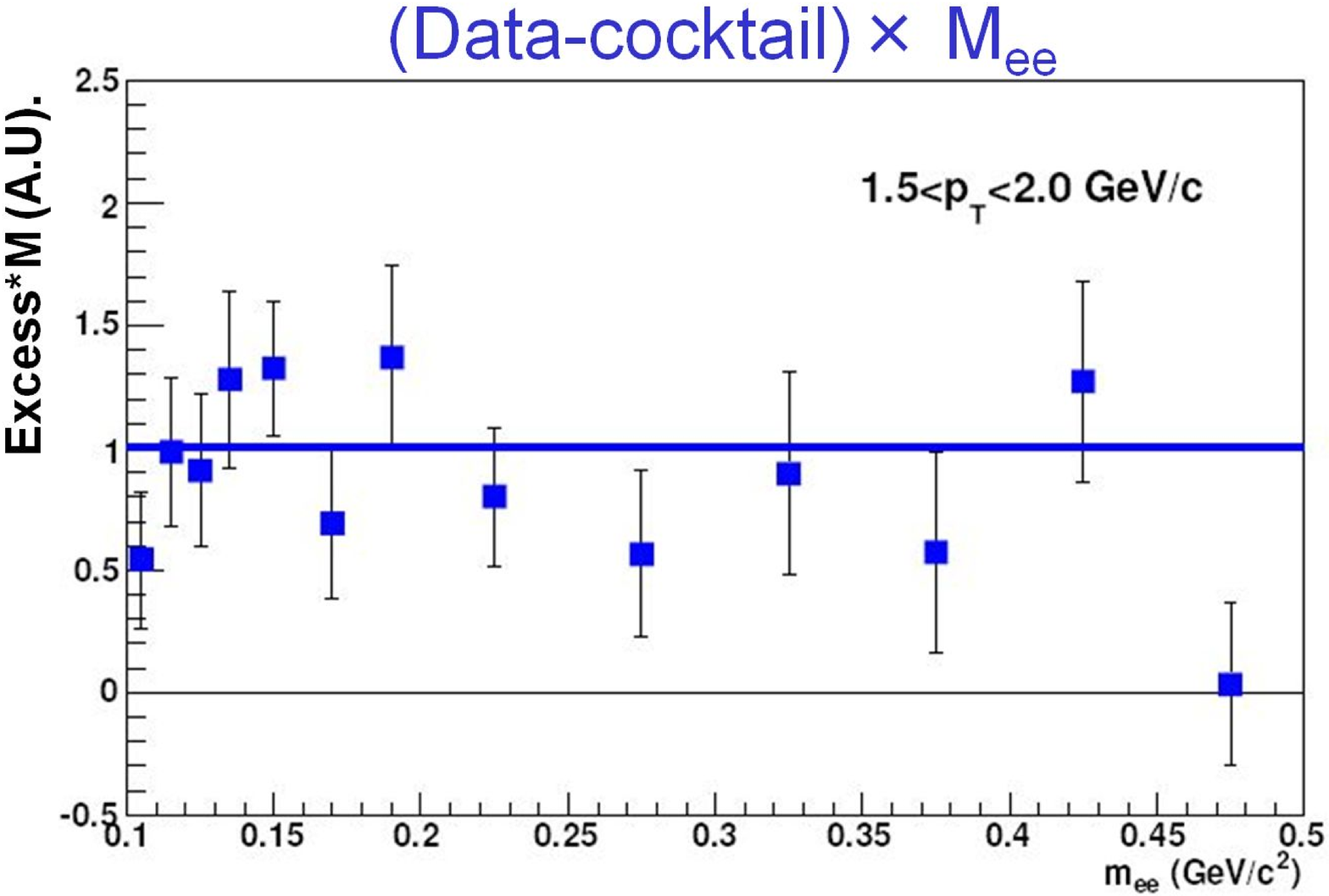}
\caption[]{
\label{fig3}
Left: Mass distributions of $e^+e^-$ pair in $p+p$ (filled) and Au+Au (open)
for $1.5<p_T<8$ GeV/$c$.
Right: $M \times dN_{ee}/dM$ of the excess $e^+e^-$ pair yield over the cocktail
for $1.5<p_T<2$ GeV/$c$ (in arbitrary unit).
}
\end{figure}

The left panel of Fig.~\ref{fig3} shows the $e^+e^-$ mass distribution for
$p_T>1.5$ GeV/$c$. The enhancement in Au+Au is still visible for $M>0.1$ GeV/$c^2$ but
its magnitude is much reduced. In the left panel, $M \times dN_{ee}/dM$ of the excess
yield of the $e^+e^-$ pair in the
Au+Au for $1.5<p_T<2.0$ GeV/$c$ is shown. Following Eq.\ref{eq:virtual} this corresponds
to the yield of virtual photons.
The figure shows that the yield of virtual photon is approximately constant
for $M>0.1$ GeV/$c$, which is consistent with the constant behavior of the HMT
calculation by R.~Rapp shown in Fig.~\ref{fig:RalfRapp}.
This could be interpreted that the virtual photon emission in the high $p_T$ region 
($p_T>1.5$ GeV/$c$)is dominated by hadronic scattering process like 
$\pi+\rho \rightarrow \pi+\gamma^*$ or partonic processes like $q+g \rightarrow q+\gamma^*$.
Both of these processes give a constant contribution to the EM spectral function at low
mass. Since the virtual photon yield is almost constant, it can be reliably
extrapolated to $M=0$. This should give the real photon emission rate.

In contrast, the mass distribution in Au+Au at low $p_T$ shown in the right panel of
Fig.~\ref{fig2} is quite different. The shape of the excess seems to be incompatible with
a constant virtual photon emission rate. The data might suggest that a large enhancement
of the EM spectral function at low mass and low $p_T$ in Au+Au collisions.


\section*{Acknowledgments} 
The author thanks R.~Rapp for very useful discussion and for providing the
calculation.

\end{document}